\documentclass[pra,twocolumn,floatfix]{revtex4-2}
\usepackage{graphicx}
\usepackage{dcolumn}
\usepackage{bm}
\usepackage{amsmath}
\usepackage{braket}
\usepackage{booktabs}
\usepackage{longtable}
\usepackage{lipsum}
\usepackage[separate-uncertainty=true,multi-part-units=single]{siunitx}
\usepackage{hyperref}
\begin{document}
\preprint{APS/123-QED}
\title{High-resolution spectroscopy and single-photon Rydberg excitation of reconfigurable ytterbium atom tweezer arrays utilizing a metastable state}
\author{Daichi Okuno}
\email{okuno.daichi@yagura.scphys.kyoto-u.ac.jp}
\author{Yuma Nakamura}
\author{Toshi Kusano}
\author{Yosuke Takasu}
\author{Nobuyuki Takei}
\altaffiliation[present address: ]{Institute of Innovative Research, Tokyo Institute of Technology, Kanagawa 226-8503, Japan}
\author{Hideki Konishi}
\author{Yoshiro Takahashi}
\affiliation{Department of Physics, Graduate School of Science, Kyoto University, Kyoto 606-8502, Japan}
\date{\today}
\begin{abstract}
    We present an experimental system for Rydberg tweezer arrays with ytterbium (Yb) atoms featuring internal state manipulation between the ground ${}^1$S$_0$ and the metastable ${}^3$P$_2$ states, and single-photon excitation from the ${}^3$P$_2$ to Rydberg states. In the experiments, single Yb atoms are trapped in two-dimensional arrays of optical tweezers and are detected by fluorescence imaging with the intercombination ${}^1$S$_0 \leftrightarrow {}^3$P$_1$ transition, and the defect-free single atom arrays are prepared by the rearrangement with the feedaback. We successfully perform high-resolution ${}^1$S$_0\leftrightarrow {}^3$P$_2$ state spectroscopy for the single atoms, demonstrating the utilities of this ultranarrow transition. We further perform single-photon excitation from the ${}^3$P$_2$ to Rydberg states for the single atoms, which is a key for the efficient Rydberg excitation. We also perform a systematic measurement of a complex energy structure of a series of D states including newly observed ${}^3$D$_3$ states. The developed system shows feasibility of future experiments towards quantum simulations and computations using single Yb atoms.
\end{abstract}
\maketitle
\section{Introduction}
\label{sec:introduction}   
    Individually trapped neutral atoms in optical tweezer arrays offer a promising platform for quantum simulation and computation~\cite{Saffman:2010, Weiss:2017, Browaeys:2020}. 
    Combination of high flexibility of the trap geometry and high tunability of interactions using Rydberg states enables a variety of experiments, ranging from observation of quantum phase transitions~\cite{Ebadhi:2021, Scholl:2020} to gate operations on multiple qubits~\cite{Levine:2018, Levine:2019, Bluvstein:2021, Graham:2021}.
    
    Two-electron atoms, or alkaline-earth-like atoms, such as Sr and Yb have several advantages over alkali atoms widely used in single atom array experiments, and some unique features have been demonstrated such as trapping of Rydberg states~\cite{Wilson:2022} and the utility of the autoionization process~\cite{Madjarov:2020,Burgers:2021}. 
    An important feature is the absence of the electron spins in the ${}^1$S$_0$ and ${}^3$P$_0$ electronic states, which makes it possible to realize nuclear spin qubits with fermionic isotopes robust against environmental magnetic field fluctuations~\cite{Gorshkov:2009}. The advantage of the use of the nuclear spin degree of freedom of ${}^{171}$Yb, in particular, in quantum information processing was discussed in~\cite{Shibata:2009, Takano:2010, Takei:2010}. Coherent nuclear spin control between two out of ten spin components in ${}^{87}$Sr~\cite{Barnes:2021} and between the two spin components in ${}^{171}$Yb~\cite{Shuo:2021, Jenkins:2021} in optical tweezers were experimentally demonstrated recently.
    
    The existence of the metastable ${}^3$P$_0$ and ${}^3$P$_2$ states is also an important property of the two-electron atoms. Quantum gate operations~\cite{Madjarov:2020, Schine:2021} and quantum metrology~\cite{Norcia:2019,Madjarov:2019,Young:2020} in tweezer array systems using the clock ${}^3$P$_0$ state have been demonstrated, while there are little investigations with the ${}^3$P$_2$ state in spite of its comparable lifetime and linewidth to those of the ${}^3$P$_0$ state. There are also several unique features of the ${}^3$P$_2$ state compared to ${}^3$P$_0$~\cite{Shibata:2009, Yamaguchi:2010}: First, the polarizability of the trapping light for the ${}^3$P$_2$ state can be widely tuned by the magnetic field orientation and thus the magic-like condition is achievable without choosing any special wavelengths. Second, the Zeeman shift of the $J=2$ electronic spin enables individual addressing of the atoms in different tweezer sites with a modest strength of the magnetic field gradient, as demonstrated for the atoms in an optical lattice~\cite{Kato:2012}. Third, the $m_J=0$ state of the bosonic isotopes is expected to be less sensitive to the magnetic field fluctuations than the ${}^3$P$_0$ state as the second-order Zeeman coefficients for the ${}^3$P$_0$ and ${}^3$P$_2$ states are caluclated to be \SI{-6.0}{Hz/G^2} and \SI{1.2}{Hz/G^2}, respectively~\cite{Dzuba:2018}. This implies that the ${}^1$S$_0$--${}^3$P$_2$ $(m_J=0)$ system of the bosonic isotopes can be a promising qubit as well as a good resource for precision measurements.

    The metastable states also offer a possibility of coherent, loss-less excitation to Rydberg states by single-photon process. Rydberg state excitation of alkali atoms is usually performed by two-photon process with an intermediate state with a short lifetime. Spontaneous decays from the intermediate state causes short coherence time even with a large detuning. Two-electron atoms, on the other hand, can be excited to Rydberg states with single-photon process from the metastable states. Coherent driving between the ${}^3$P$_0$ and a Rydberg state was indeed demonstrated with ${}^{88}$Sr~\cite{Madjarov:2020}.
    
    In this paper, we present an experimental system for Rydberg atom arrays with Yb atoms featuring internal state manipulation between the ground ${}^1$S$_0$ and the metastable ${}^3$P$_2$ states, and single-photon excitation from the ${}^3$P$_2$ to Rydberg states. We trap single ${}^{174}$Yb atoms in arrays of optical tweezers generated by a pair of acousto-optic deflectors (AODs). High-sensitive imaging of trapped single atoms is performed using the narrow-line ${}^1$S$_0\leftrightarrow{}^3$P$_1$ transition, yielding 95\% of imaging fidelity. The trapped single atoms in one dimension (1D) and two dimension (2D) are dynamically rearranged to defect-free single atom arrays with a fast feedback system. Using this system, we succeed in high-resolution ${}^1$S$_0\leftrightarrow{}^3$P$_2$ laser spectroscopy for single Yb atoms. We also successfully demonstrate single-photon excitation to a Rydberg state from the metastable ${}^3$P$_2$ state for single atoms in optical tweezers. In addition, we perform a systematic laser spectroscopy for Rydberg states using evaporatively cooled atoms, in which we newly observe a series of (6s)($n$d)${}^3$D$_3$ states ranging $n=65$--$80$. This work offers an important step towards the realization of scalable systems for versatile quantum simulation and computation applications using Yb atom tweezer arrays.
	
	This paper is structured as follows: We present in section~\ref{sec:apparatus} the overview of the apparatus including the vacuum chamber system integrating electrodes and a micro-channel plate (MCP), an optical tweezer array generation system and laser sources for internal state control and probing. In section~\ref{sec:imaging}, we show the result of single atom trapping and imaging in optical tweezer arrays, as well as tweezer array rearrangement by a feedback program. In section~\ref{sec:3p2}, we describe the ${}^1$S$_0\leftrightarrow{}^3$P$_2$ laser spectroscopy for single atoms in an optical tweezer array. In section~\ref{sec:rydberg}, we show the result of single-photon Rydberg state spectroscopy from the $^3$P$_2$ state with both single atoms and an ensemble of evaporatively cooled atoms.
	Finally in section~\ref{sec:conclusion}, we conclude our work with perspectives using the developed system towards quantum simulation and computation.
	
        \begin{figure}[htbp]
        	\centering
        	\includegraphics[width=\linewidth]{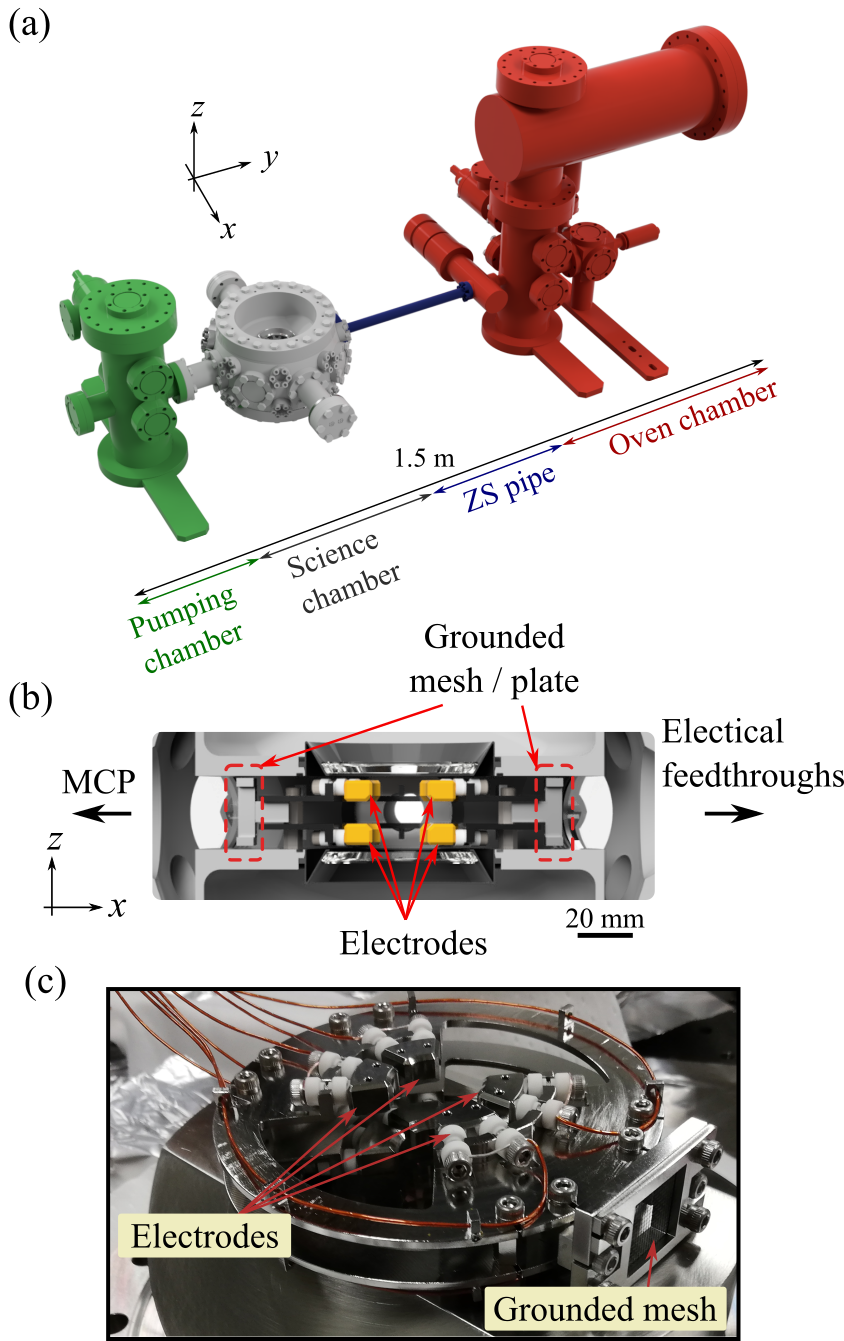}
        	\caption{\label{fig:overview}{
        	(a)~Overview of the whole vacuum chamber consisting of oven chamber (red), Zeeman slower pipe (blue), science chamber (grey), and pumping chamber (green).
        	(b)~Vertical cut view of the science chamber in the $x$-$z$ plane. Eight electrodes (yellow) are held on the bottom viewport. The MCP and electrical feedthroughs are attached to the chamber. A grounded mesh and an aluminum plate placed next to the electrodes shield atoms from electric fields created by the MCP and feedthroughs, respectively.
        	(c)~Photograph of the electrodes and the mesh assembled on the bottom viewport prior to installation in the vacuum system.
        	}}
        \end{figure}
        
\section{Experimental Apparatus}
\label{sec:apparatus}
    \subsection{Vacuum chamber}
	        
        Our vacuum chamber system is based on a rather standard system for producing Yb quantum gases, and it comprises four sections: oven chamber, Zeeman slower pipe, science chamber, and pumping chamber, as depicted in Fig.~\ref{fig:overview}(a). In particular, the science chamber is specially designed for experiments of the Rydberg atom detection and optical tweezer arrays, based on a commercial chamber (Kimball Physics, MCF800-ExtOct-G2C8A16) with two ICF203, eight ICF70 and sixteen ICF34 flanges. Two reentrant viewports are mounted on the ICF203 flanges on top and bottom, hosting high-NA objective lenses for high-resolution imaging, optical tweezer arrays, and site-resolved manipulation. Electrical feedthroughs and a MCP (Hamamatsu photonics F4655-11) for Rydberg state detection are attached to the chamber. Many viewports allow optical access for laser cooling, internal state excitation, absorption and fluorescence imaging, and further optical trapping.

        In addition, the science chamber is equipped with eight electrodes as shown in Fig.~\ref{fig:overview}(b) and (c) in order to compensate for stray electric fields at the position of the atoms as well as to ionize Rydberg atoms, enabling high-sensitive detection of Rydberg states by the MCP. The eletrodes are connected to a digital-analog converter through the feedthroughs, which outputs $\pm\SI{1}{V}$ with 16-bit resolution, corresponding to \SI{0.9}{\micro V/mm}, \SI{1.2}{\micro V/mm}, and \SI{0.7}{\micro V/mm} resolution along the $x$, $y$, and $z$ directions, respectively. 
        The atoms are shielded from electric fields created by the MCP and the feedthroughs with a grounded mesh on the MCP side and an aluminum plate on the feedthrough side.
        In order to avoid charging effects on the dielectric coating, the reentrant viewports are processed with indium tin oxide and grounded. 
        
    \subsection{Laser sources}  
        \begin{figure}[ht]
            \centering
            \includegraphics[width=.9\linewidth]{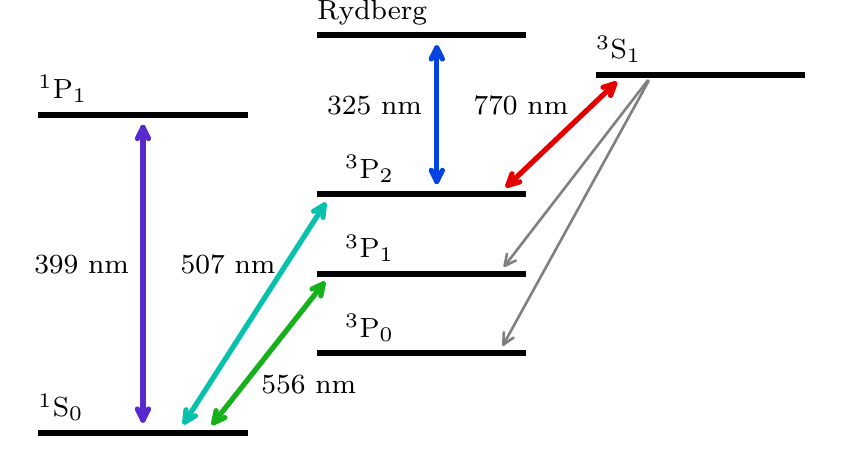}
            \caption{Energy levels of Yb relevant for the experiment. The strong ${}^1$S$_0\leftrightarrow {}^1$P$_1$ transition (\SI{399}{nm}) is used for the Zeeman slower and absorption imaging. The narrow-line ${}^1$S$_0 \leftrightarrow {}^3$P$_1$ transition (\SI{556}{nm}) is used for the MOT and fluorescence imaging for single atoms. Atoms in the metastable ${}^3$P$_2$ state are prepared by direct excitation from the ground state with a \SI{507}{nm} light. The atoms in the ${}^3$P$_2$ state are excited to Rydberg states by single-photon excitation with a \SI{325}{nm} laser. For high-resolution spectroscopy of the ${}^1$S$_0\leftrightarrow {}^3$P$_2$ transition, the ${}^3$P$_2$ atoms are repumped to the ground state through the ${}^3$S$_1\rightarrow{}^3$P$_1\rightarrow{}^1$S$_0$ process by applying a \SI{770}{nm} laser.}
            \label{fig:levels}
        \end{figure}
        
        We utilize five optical transitions shown in Fig.~\ref{fig:levels}.
        The Zeeman slowing beam operating on the $^1$S$_0\leftrightarrow ^1$P$_1$ transition with a wavelength of \SI{399}{nm} is generated by a second harmonic generation (SHG) of a \SI{798}{nm} laser beam derived from a tapered-amplified diode laser (Toptica, TA-pro), the frequency of which is stabilized with a wavemeter (High Finesse, WS-8 10). 
        The magneto-optical trapping (MOT) and fluorescence imaging are operated on the $^1$S$_0\leftrightarrow ^3$P$_1$ transition with a wavelength of \SI{556}{nm}. They are also derived from a SHG of a tapered-amplified diode laser (Toptica, TA-pro) with a wavelength of \SI{1112}{nm}. The diode laser is frequency-stabilized with the Pound--Drever--Hall (PDH) technique to an ultra-low expansion (ULE) glass cavity with a finesse of about 20,000 at this wavelength.
        
        Towards a qubit operation and subsequent Rydberg excitation, the atoms in the ground state ${}^1$S$_0$ are directly excited to the ${}^3$P$_2$ state with a resonant \SI{507}{nm} light. We generate the \SI{507}{nm} light from a SHG of a home-built external cavity diode laser (ECDL)--tapered amplifier (TA) system at \SI{1014}{nm}. We obtain \SI{15}{mW}, which is enough for us to perform a Rabi oscillation at about several kHz. The frequency of the ECDL is also stabilized to the same ULE glass cavity with a finesse of about 500,000 at this wavelength. The linewidth is estiamted to be \SI{200}{Hz} from the PDH error signal at \SI{1014}{nm}. The \SI{507}{nm} beam propagates along the radial direction of the optical tweezers and has a horizontally elongated shape of $\SI{60}{\micro \metre} \times \SI{530}{\micro \metre}$ at the position of the atoms to ensure uniform intensity over the array.
        In the $^1$S$_0\leftrightarrow ^3$P$_2$ spectroscopy, we apply repumping light resonant with the $^3$P$_2\leftrightarrow\ ^3$S$_1$ transition with a wavelength of \SI{770}{nm} for the ${}^3$P$_2$ atoms to return to the ground state via the ${}^3$P$_1$ state. The light is derived from an interference filter diode laser, frequency-stabilized with the wavemeter.
        The atoms in the ${}^3$P$_2$ state are transferred to Rydberg states by a \SI{325}{nm} light generated by a SHG of a home-built ECDL--TA system at with a wavelength of \SI{650}{nm}, the frequency of which is stabilized with the wavemeter. 
        We obtain \SI{30}{mW} with this system.

    \subsection{Optical tweezer array system}
    
\begin{figure}[t]
    	    \centering
        	\includegraphics[width=\linewidth]{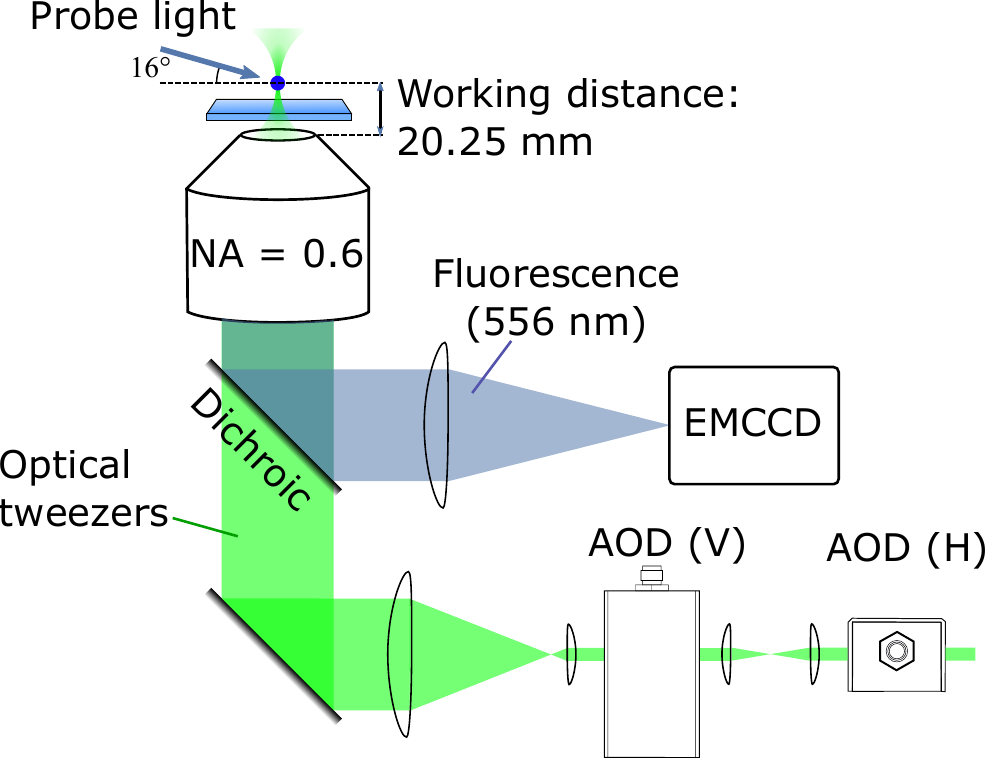}
        	\caption{\label{fig:Tweezer-schematic}{
        	Sketch of the optical system for the optical tweezers and fluorescence imaging. Two AODs are placed horizontally (AOD(H)) and vertically (AOD(V)) on the imaging planes of the standard $4f$ configuration to generate two-dimensional tweezer arrays. The deflected \SI{532}{nm} beams pass through the dichroic mirror and the objective lens with NA = 0.6 and focused onto the atoms. \SI{556}{nm} fluorescence from the atoms are collected on the EMCCD camera through the objective lens.
        	}}
        \end{figure}

        The optical tweezers with a wavelength of \SI{532}{nm} are derived from a diode-pumped solid-state laser (Coherent, Verdi V-10). The laser beam is arranged into two-dimensional arrays of optical tweezers by two orthogonally oriented AODs (AA opto-electronic, DTSX-400-532) in the $4f$ configuration (see Fig.~\ref{fig:Tweezer-schematic}). We apply multi-tone radio-frequencies (RFs) generated by an arbitrary waveform generator (Spectrum, M4i.6622-x8) to the AODs. Relative phases of the RF components are designed to reduce the peak-to-average power ratio using the Kitayoshi algorithm~\cite{Kitayoshi:1985}.

        The deflected beams are magnified and pass through an objective lens below the science chamber. 
        The objective lens has an NA of 0.6 and a working distance of \SI{20.25}{mm} and it is designed to be diffraction-limited for \SI{532}{nm} and \SI{556}{nm} light, corresponding to the tweezer and imaging wavelengths, respectively.
        The tweezers are focused down to a waist of \SI{0.89}{\micro m}, determined by the sideband-resolved ${}^1$S$_0\leftrightarrow{}^3$P$_2$ spectroscopy as described in section~\ref{sec:3p2}.
        The deflection angle of the AOD is \SI{0.8}{mrad/MHz}, resulting in a \SI{2.4}{\micro m/MHz} shift at the position of the atoms. 
       
\section{Single atom trapping and imaging}
\label{sec:imaging}
    The experiment starts from loading atoms in a MOT from a Zeeman-slowed atomic beam for \SI{500}{ms}.
    The atoms collected in the MOT are then compressed by increasing the magnetic field gradient and further cooled by decreasing the detuning and the power of the MOT beams.
    Typical temperature and the number of atoms at this stage are $\sim \SI{20}{\micro K}$ and $10^5$, respectively.
    Optical tweezers are overlapped with the MOT cloud, and some of the atoms are trapped in the optical tweezers after turning off the MOT beams.
    \begin{figure}[ht]
	    \centering
    	\includegraphics[width=.95\linewidth]{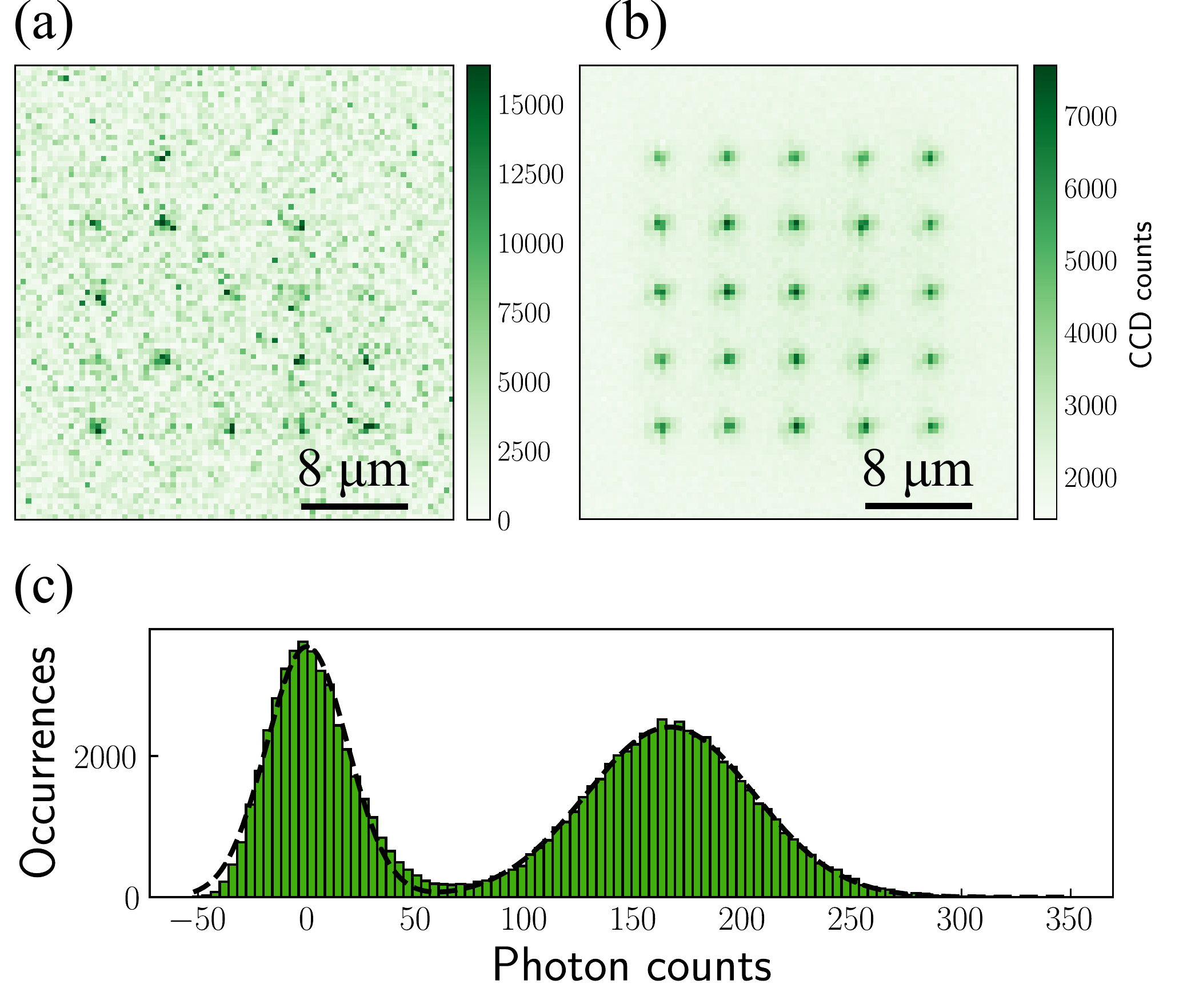}
    	\caption{\label{fig:single-atoms}{
    	(a)~Single-shot and (b)~400-average image of atoms in a $5\times 5$ tweezer array with the spacing \SI{4.8}{\micro m} (\SI{60}{ms} exposure time). 
    	(c)~Histogram of photons detected by the CCD camera in each $3\times 3$ pixel${}^2$ region around the atom positions. The dashed line represents the fitted curve by a double gaussian distribution. The dark count is set to be zero for clarity.
    	}}
    \end{figure}
    Typical power of each tweezer is \SI{18}{mW}, corresponding to the potential depth of $k_\mathrm{B} \times \SI{660}{\micro K}$, where  $k_\mathrm{B}$ is the Boltzmann constant.
    We apply a red-detuned \SI{556}{nm} light to induce losses of pairs of atoms by light-assisted collisions, yielding zero or one atom in each tweezer reflecting the parity of the initial number of atoms~\cite{Fuhrmanek:2012}. 
    
    Single atoms are detected with fluorescence imaging with the ${}^1$S$_0 \leftrightarrow {}^3$P$_1$ transition. The imaging light is tilted by \SI{16}{\degree} from the horizontal plane and horizontally polarized.
    Photons emitted from the atoms are collected for \SI{60}{ms} on a charge-coupled device camera (Andor, iXon-885) through the objective lens  (Fig.~\ref{fig:Tweezer-schematic}). 150 photons per single atom are typically detected. 
 
    Single shot and 400-averaged images of a $5\times 5$ tweezer array with a \SI{4.8}{\micro m} spacing are shown in Fig.~\ref{fig:single-atoms}(a) and (b), respectively. 
    A binarized behavior of the fluorescent photon counts is clearly observed in the histogram as shown in Fig.~\ref{fig:single-atoms}(c). 
    We determine the count threshold to distinguish between filled and empty sites by a double Gaussian fit. 
    We achieve a false undetection and a false detection error probability of 0.11\% and 0.24\%, respectively.
    
\begin{figure}[ht]
	    \centering
    	\includegraphics[width=.95\linewidth]{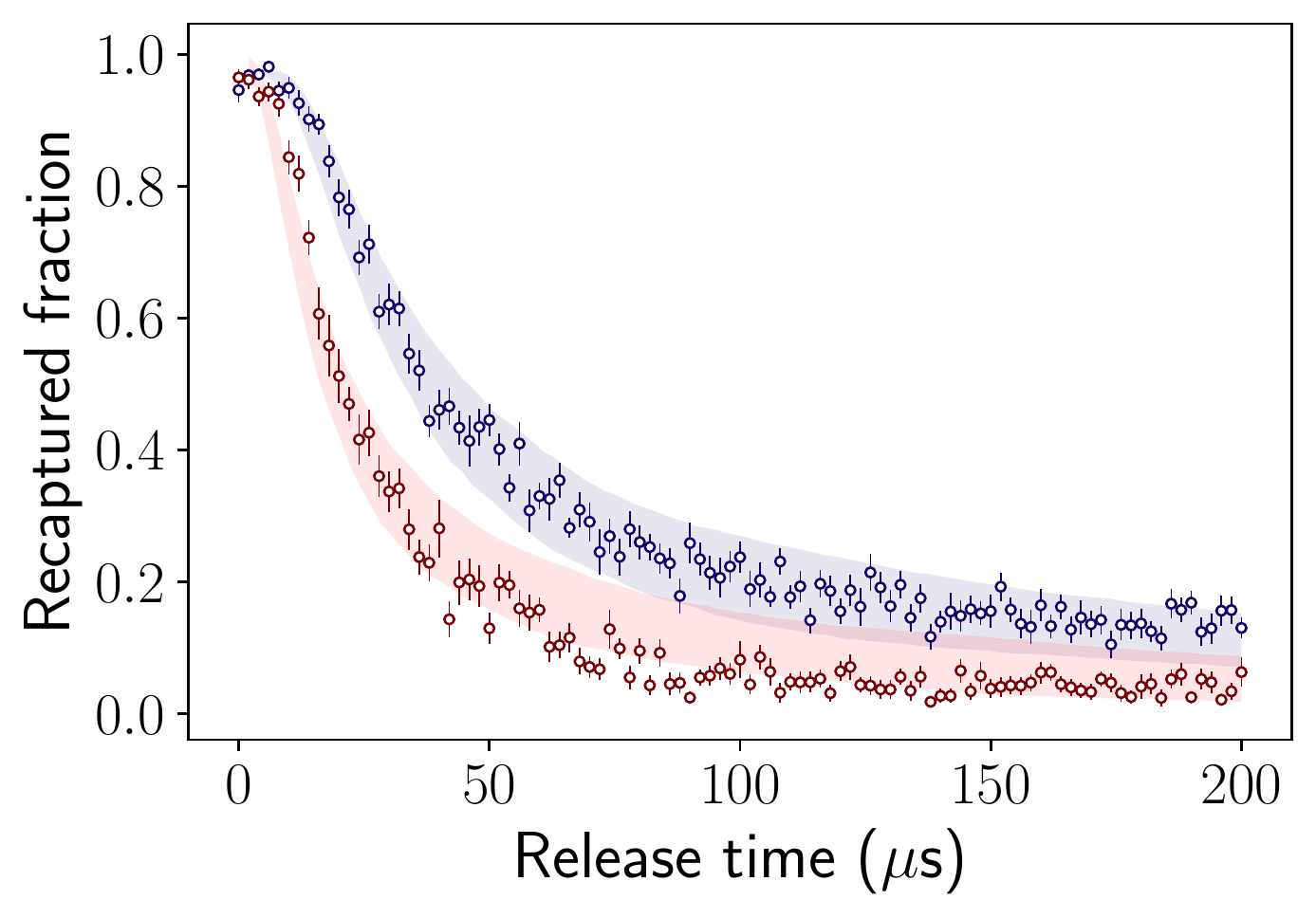}
    	\caption{\label{fig:RandR}{Recapture probability as a function of the release time $t$ with (blue) and without (red) the cooling process.
    	The error bars show the standard error of 16 repetitions.
    	The shaded area indicates a 95\% confidential interval from 100 trials of the Monte-Carlo simulation at the best fit temperature $T=\SI{19.5}{\micro K}$ (blue) and $T=\SI{59.1}{\micro K}$ (red).
    	}}
    \end{figure}

    Atoms heated by the probe light are cooled after the image using the ${}^1$S$_0\leftrightarrow {}^3$P$_1$ MOT lights in three directions for \SI{10}{ms}, which is red-detuned by 1.9 times the natural linewidth.
    The cooling effect is verified by the release-and-recapture technique~\cite{Tuchendler:2008}. We first take an image of an atom array followed by the cooling. We then turn off the trap for a release time $t$ and take the second image. We repeat this procedure 16 times for each $t$ and obtain the probability of recapturing atoms as a function of $t$ with and without the cooling as shown in Fig.~\ref{fig:RandR}.
    We fit the experimental data to Monte-Carlo simulated trajectories by the weighted least-square method 100 times and we deduce the temperature to be \SI{19.5\pm2.1}{\micro K} for the data with cooling and \SI{59.1\pm5.4}{\micro K} for without cooling, from the mean and the standard deviation of the fit results. The temperature with cooling is close to that in the MOT. We note that the obtained temperature corresponds to that in the radial direction since the radial expansion is critical for the atoms to remain in the highly anisotropic recapture region.
    
\begin{figure}[ht]
	    \centering
    	\includegraphics[width=.95\linewidth]{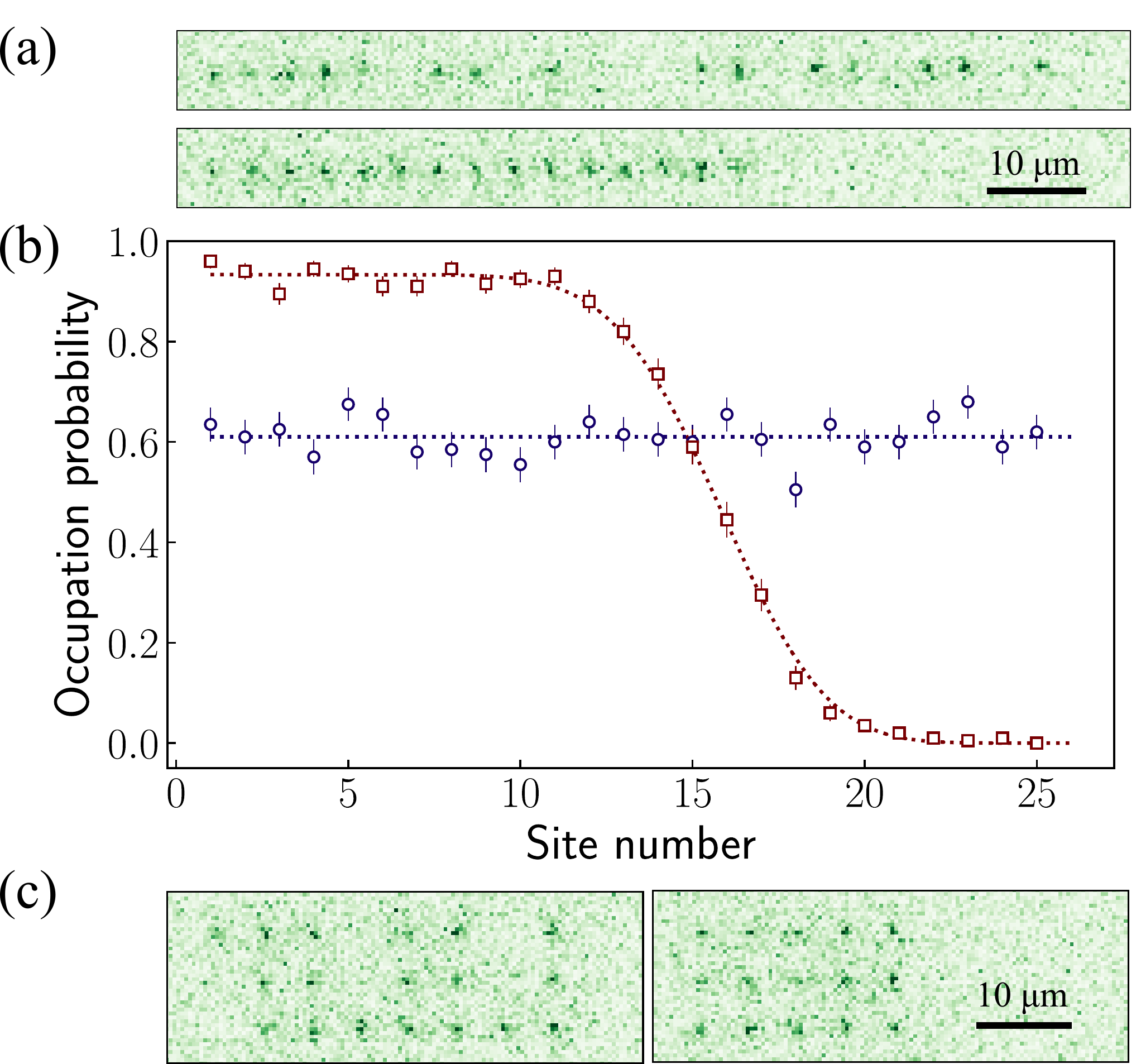}
    	\caption{\label{fig:rearrange}{(a)~Fluorescence images before (upper) and after (lower) the rearrangement of atoms for one-dimensional 25 tweezer array. (b)~The occupation probability of each site before (blue circles) and after (red squares) the rearrangement averaged for 200-shots. (c)~Fluorescence images before (left) and after (right) the rearrangement of the atoms for a two-dimensional $3\times 8$ tweezer array. Only defectless columns are left and shifted to form a defect-free array.
    	}}
    \end{figure}

    To obtain defect-free atom arrays~\cite{Endres:2016,Daniel:2016}, we apply a rearrangement protocol in 1D and 2D arrays. We first prepare a sample randomly loaded in a 1D array of 25 sites and take an image. At this stage, we identify the filled and empty sites.
    We then turn off the RF components applied to the AODs corresponding to the empty trap sites, and move the remaining sites to the left by dynamically changing the multi-tone RF in \SI{0.64}{ms}.
    To confirm the successful rearrangement, we finally take the second image.
    Figure~\ref{fig:rearrange}(a) shows typical images before and after the rearrangement.
    
    Figure~\ref{fig:rearrange}(b) shows the probability to find an atom in each site obtained by 200 experimental runs before (blue circles) and after (red squares) the rearrangement.
    The probability to find an atom in the $i$-th site from the left edge after the rearrangement is given by
    \begin{eqnarray}
       P_i = (1-q) \times \sum_{\substack{j\\ {i\leq j \leq N}}} \binom{N}{j}\ p^j\ (1-p)^{N-j}. \label{eq:rearrange}
    \end{eqnarray}
    Here $N$ is the number of total trap sites before the rearrangement, $p$ is the loading probability, and $q$ is the loss probability between the two images. 
    The fit of Eq.~(\ref{eq:rearrange}) gives $p = 0.61$ and $q = 0.07 $.
    The obtained value of $p$ and $1-q$ are close to the fully random case of 1/2, and the survival probability of 0.95, which indicates the rearrangement process is performed with negligibly small error.  

    This rearrangement method using the AODs is not limited to 1D arrays. Figure \ref{fig:rearrange}(c) shows an example for a 2D array. By only collecting defectless vertical lines of sites in a $3\times 8$ array, we successfully prepare a defect-free $3\times 5$ array of single atoms. 
    As far as we know, this is the first demonstration of creating a defect-free 2D array only by AODs.  
\section{Excitation of the single atoms to the metastable $^3$P$_2$ state}
    \label{sec:3p2}
    We perform high-resolution laser spectroscopy for the  transition between the ${}^1\mathrm{S}_0$ and ${}^3\mathrm{P}_2$ states of single atoms in an optical tweezer array. 
    Unlike the ${}^1$S$_0\leftrightarrow {}^3$P$_0$ transition, the differential light shift between the ${}^1\mathrm{S}_0$ and ${}^3\mathrm{P}_2$ states can be controlled by tuning the angle between the polarization of the tweezers and the magnetic field direction. For the ${}^{174}$Yb isotope in \SI{532}{nm} light, the difference of the light shifts between ${}^3$P$_2$ ($m_J=0$) and ${}^1$S$_0$ is minimized such that the ratio of their polarizabilities is 1.016 when the magnetic field is aligned to the light polarization~\cite{Tomita:2019}.
    We apply \SI{2.8}{G} in order to lift the degeneracy along the direction of tweezer light polarization. 
    
\begin{figure}[ht]
	    \centering
    	\includegraphics[width=.95\linewidth]{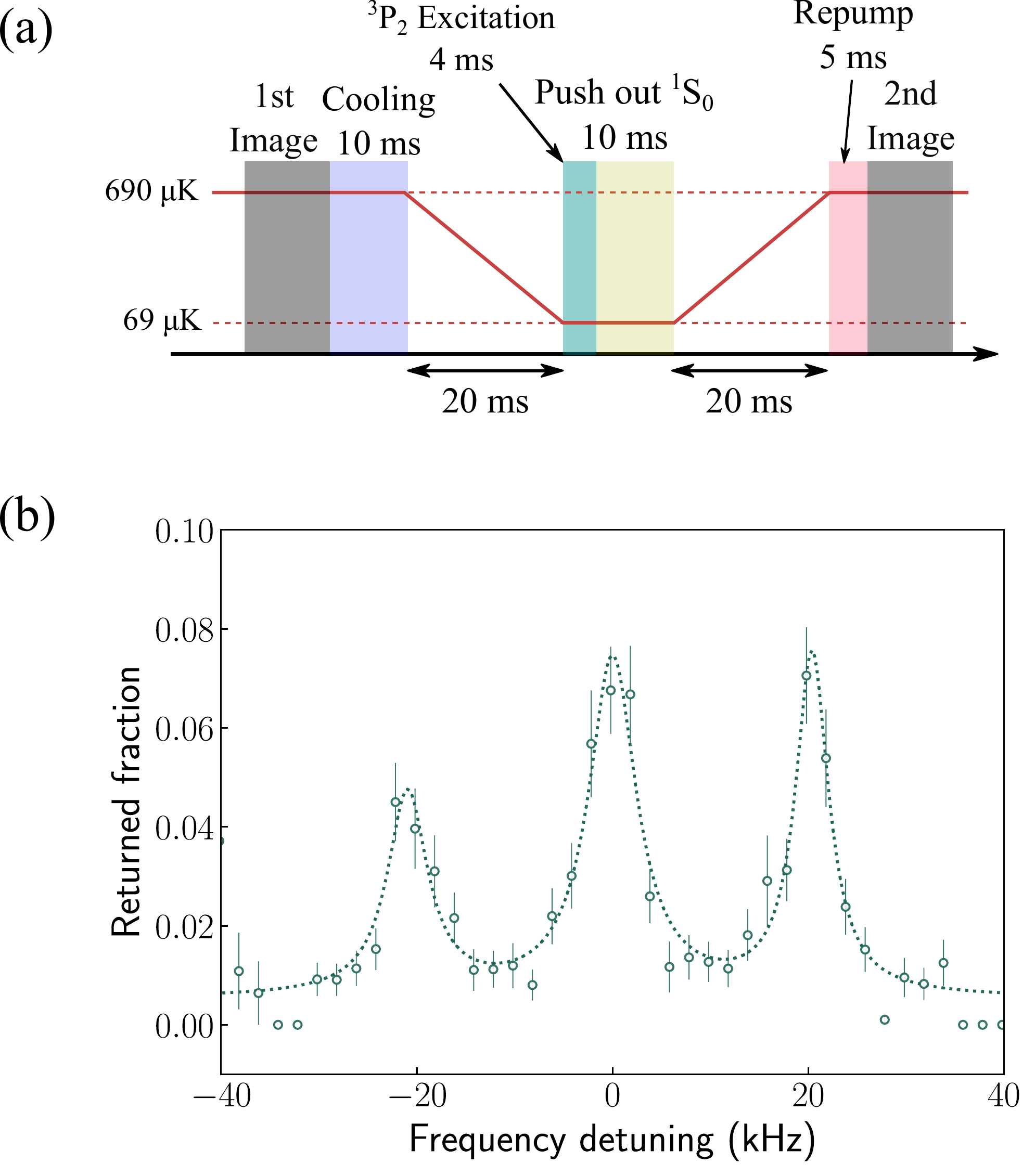}
    	\caption{\label{fig:3P2-spectrum}{
    	(a)~Pulse sequence for ${}^3$P$_2$ state spectroscopy.
    	(b)~A ${}^3$P$_2\ (m_J = 0)$ spectrum for single atoms in the $5\times 5$ tweezer array. The dotted line is a fit by three Lorentzian functions. The peak height ratio of the red to the blue sidebands yields the temperature \SI{1.90\pm 1.02}{\micro K} and the mean vibrational number occupation $n=1.46\pm 1.01$.
    	}}
    \end{figure}

    We perform spectroscopy of the $m_J=0$ state of single atoms in a $5\times 5$ array. The experimental sequence proceeds as shown in Fig.~\ref{fig:3P2-spectrum}(a). We first take an image of single atoms randomly loaded in a $5 \times 5$ array followed by the cooling as described in section~\ref{sec:imaging}. The potential depth at this stage is \SI{690}{\micro K}. The tweezer potential is adiabatically ramped down to $1/10$ to further reduce inhomogeneous broadening of the ${}^1$S$_0\leftrightarrow{}^3$P$_2$ resonance. We apply a \SI{507}{nm} laser pulse for \SI{4}{ms} followed by a \SI{556}{nm} laser pulse for \SI{10}{ms} to remove the remaining atoms in the ground state over the array. After the \SI{507}{nm} excitation, we ramp the tweezer potential back to the initial value and apply a \SI{770}{nm} light for \SI{5}{ms}. We note that the ${}^3$S$_1$ state spontaneously decays into all the ${}^3$P$_J\ (J=0,1,2)$ states but majority of the ${}^3$P$_2$ atoms return to the ${}^1$S$_0$ state through the ${}^3$P$_1$ without the other repumper ${}^3$P$_0\leftrightarrow {}^3$S$_1$~\cite{Cho:2012}. Finally, we take the second fluorescence image and obtain the return fraction by comparing with the first image.
    
    Figure~\ref{fig:3P2-spectrum}(b) shows a typical spectrum. The carrier resonance and the blue and red sidebands are well resolved, promising for the sideband cooling to the vibrational ground state in the optical tweezers.
    The radial trap frequency $\Omega_r/(2\pi)$ deduced from the sidebands is \SI{20.7\pm 0.3}{kHz}, from which we evaluate the waist of the tweezer beams to be \SI{0.89}{\micro m}.
    The ratio of the peak heights of the red sideband $A_\mathrm{R}$ to the blue $A_\mathrm{B}$ gives the temperature $T$ and the mean vibrational number occupation $\bar{n}$ by a relation $A_\mathrm{R}/A_\mathrm{B} = \exp\left(-\hbar \Omega_r / (k_\mathrm{B}T)\right)=\bar{n}/(1+\bar{n})$ where $\hbar$ denotes the reduced planck constant, and we find $T=\SI{1.90\pm 1.02}{\micro K}$ and $\bar{n}=1.46\pm 1.01$.
    The obtained value of $\bar{n}$ is in agreement with the theoretical expectation that $\bar{n}$ is reduced to $1.05$ at most by the cooling process taking into account the sideband transitions, using Eq.~(121) of \cite{Leibfried:2003}.
    The assumption of the same distribution of the occupation for trap depth before the ramping down of the potential suggests the temperature of \SI{6.01 \pm 3.24}{\micro K}. We attribute the difference between the value obtained with the release-and-recapture method in the previous section to the assumption of isotropic temperatures in the Monte-Carlo simulation. Note that the saturation effect would contribute to reduction of the peak height of the carrier resonance.
\section{Rydberg state spectroscopy}
    \label{sec:rydberg}
    \begin{figure}[ht]
	    \centering
    	\includegraphics[width=.95\linewidth]{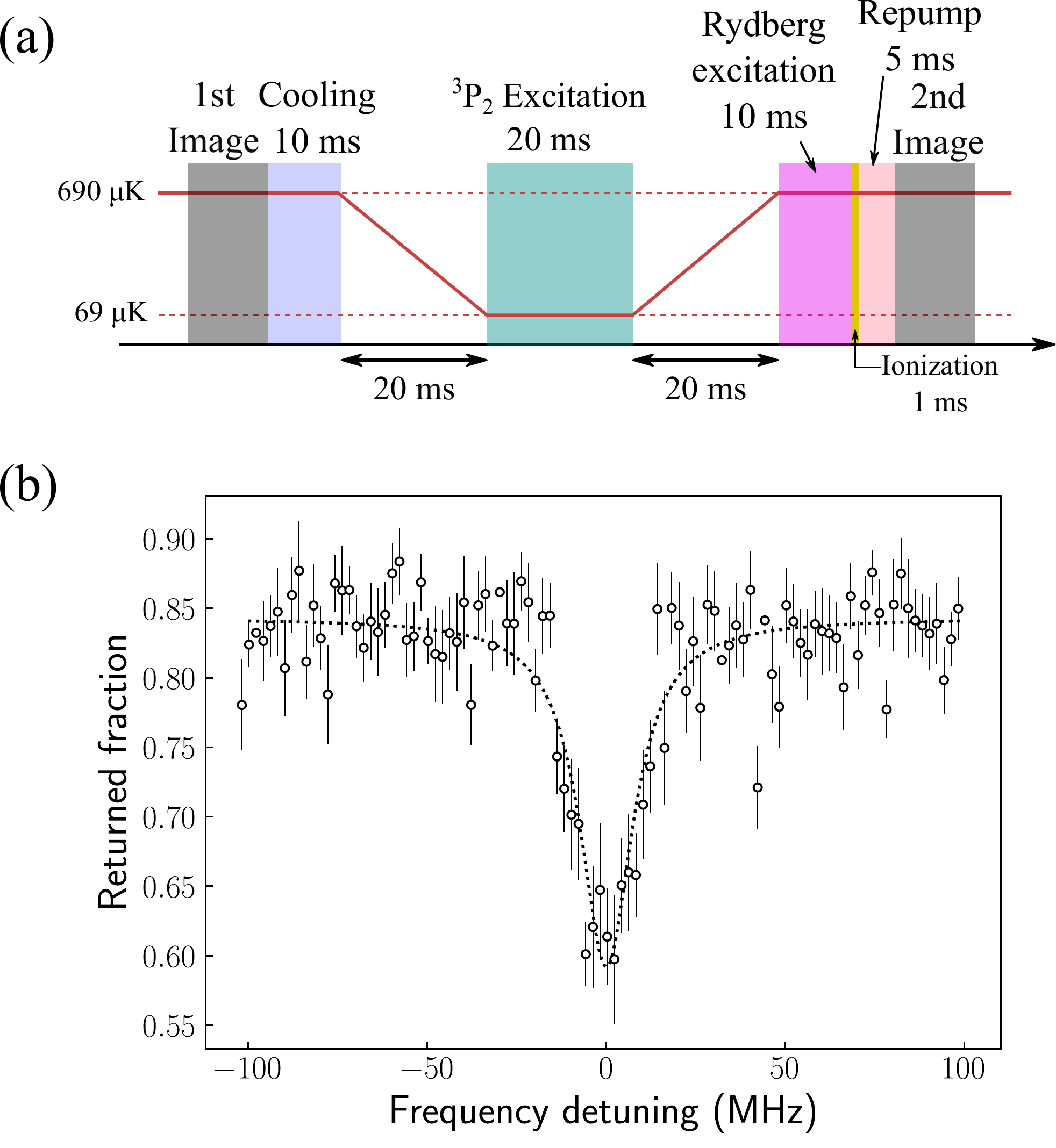}
    	\caption{\label{fig:Rydberg-tweezer}{
    	(a)~Pulse sequence for the Rydberg spectroscopy from the ${}^3$P$_2$ state for single atoms.
    	(b)~A spectrum of the ${}^3$P$_2$, $m_J=0 \leftrightarrow\ \mathrm{(6s)(78d)}{}^3$D$_3$ transition for single atoms in the $5 \times 5$ tweezer array. Rydberg states are ionized by an electric field before the repumping of ${}^3$P$_2$ state and observed as a loss of the ${}^1$S$_0$ state.
    	}}
    \end{figure}
    We perform single-photon Rydberg state excitation spectroscopy from the ${}^3$P$_2$ state in single atom arrays. The ${}^3$P$_2$ state atoms are prepared by applying a \SI{507}{nm} laser pulse for \SI{20}{ms} on the carrier resonance in the shallow trap, yielding the excitation probability of 40\%. After ramping back the trap potential we apply a \SI{325}{nm} laser pulse for \SI{10}{ms} followed by ionization. We then take the second image after a \SI{5}{ms} pulse of the \SI{770}{nm} laser (see Fig.~\ref{fig:Rydberg-tweezer}(a)). Successful excitation to a Rydberg state is observed as a decrease of the return probability as shown in Fig.~\ref{fig:Rydberg-tweezer}(b). The observed resonance corresponds to the (6s)(78d) ${}^3$D$_3$ state. This is the first demonstration of the single-photon Rydberg excitation of a single Yb atom array.
    
    \begin{figure*}[ht]
	    \centering
    	\includegraphics[width=\linewidth]{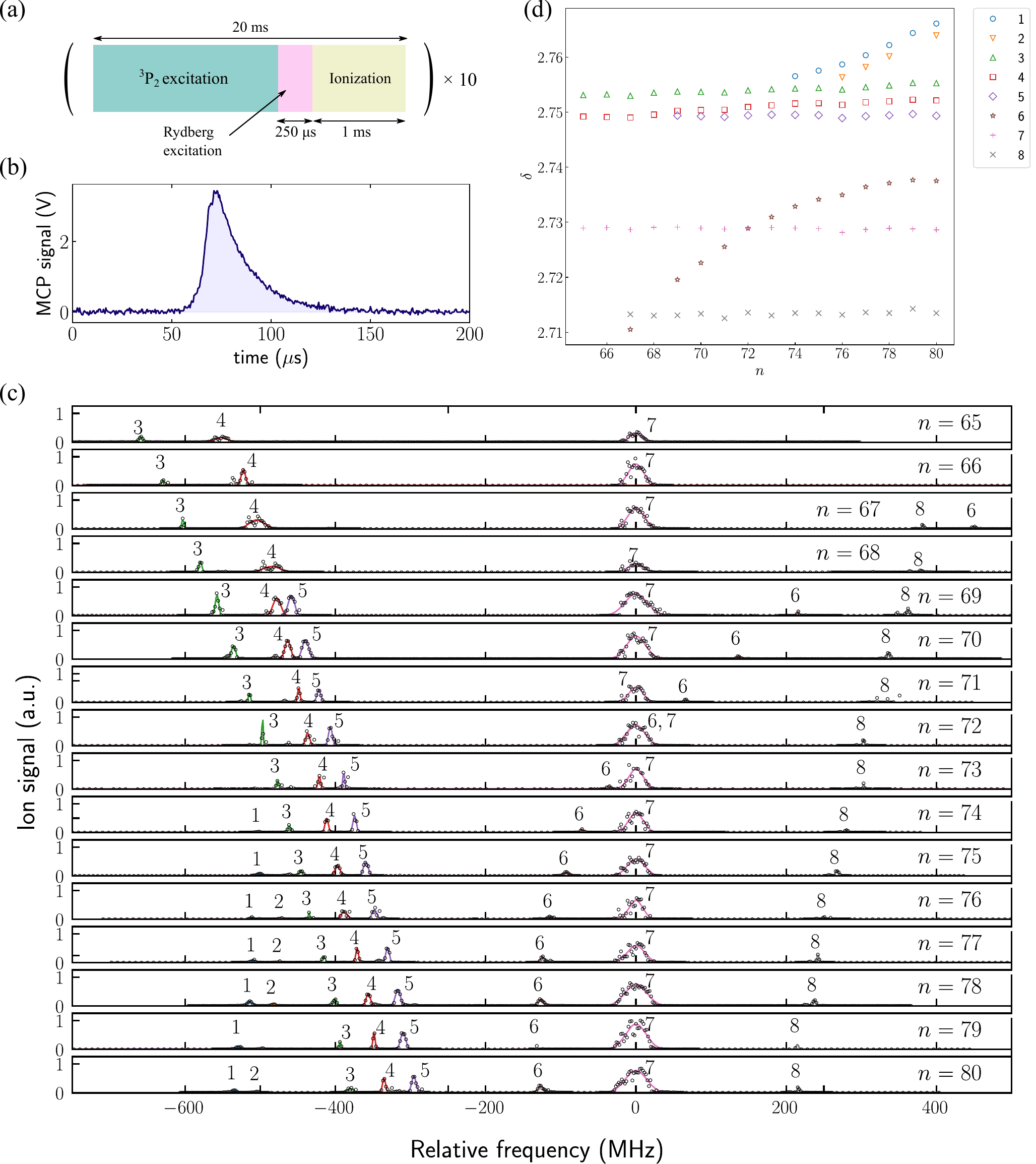}
    	\caption{\label{fig:Rydberg_spectrum}{
    	(a)~Pulse sequence for the Rydberg state ionization spectroscopy.
    	(b)~Typical MCP signal averaged for 10 pulses. The area of the trace corresponds to the number of ions detected by the MCP. 
    	(c)~Spectra around the ${}^3$P$_2\ (m_J = 2)\leftrightarrow\ {}^3$D transition for $n=65$--80. We label the resonances by numbers from lower to higher energies in the $n=80$ spectrum. The horizontal axis is relative freuqency from the strongest resonance 7. The resonances 3, 5 and 8 correspond to ${}^3$D$_1$, ${}^3$D$_2$ and ${}^1$D$_2$ states, respectively. The peaks 2 and 7 are the new observations, and 7 is assigned as (6s)($n$d)${}^3$D$_3$. Note that the peaks of 1, 4, and 6 were also observed in \cite{SaskinPhD} but not assigned.
    	(d)~The quantum defect $\delta$ of the resonances against the principal quantum number.
    	}}
    \end{figure*}
    
    In addition, the spectroscopic information on the accessible Rydberg states is crucially important, and thus we further perform systematic spectroscopic measurements of the transition between the ${}^3$P$_2$ and Rydberg states. In this measurement, different from the previous experiments, we use an ensemble of atoms in an optical dipole trap (ODT) with a wavelength of \SI{1070}{nm} for the sake of high signal intensity. 
    Atoms loaded in a MOT for \SI{10}{s} are transferred into the ODT and evaporatively cooled. Finally we obtain $10^5$ atoms with a temperatrue of \SI{5}{\micro K}. We first apply a \SI{507}{nm} pulse resonant to the ${}^1$S$_0\leftrightarrow{}^3$P$_2$ $(m_J=2)$ transition for \SI{18.75}{ms} followed by a \SI{325}{nm} pulse for \SI{250}{\micro s}.
    Immediately after the \SI{325}{nm} pulse we ramp up the electric field to \SI{17}{V/cm} to detect the Rydberg states on the MCP by the field ionization method. 
    The MCP signal is amplified and acquired by a field-programmable gated array with a \SI{2}{\micro s} time resolution. 
    This excitation and ionization process is repeated for 10 times and the MCP signals are averaged. Figure~\ref{fig:Rydberg_spectrum}(a) and (b) shows the pulse sequence and a typical averaged MCP signal, respectively. The integral of the MCP signal corresponds to the number of the ionized Rydberg state atoms.
    
    We perform spectroscopy around the D series of Rydberg states from $n=65$ to 80 as shown in Fig.~\ref{fig:Rydberg_spectrum}(c). We observe up to eight resonances in the range of $\sim \SI{1}{GHz}$. The resonances indexed as 2 and 7 are newly observed in this work, while the rest have been observed in the previous works and 3, 5 and 8 were assigned to be ${}^3$D$_1$, ${}^3$D$_2$ and ${}^1$D$_2$ states, respectively~\cite{Lehec:2018}. We consider the strongest resonance 7 to be the ${}^3$D$_3$ state because of the selection rule of the dipole transition. This assignment is supported by the fact that the resonance splits into four corresponding to $|m_J| = 0,1,2,3$ manifolds by applying electric fields due to the electric-field-induced quadruple splitting.
    
    We deduce the quantum defect of each resonance and plot them against the principal quantum number in Fig.~\ref{fig:Rydberg_spectrum}(d). Here the quantum defect $\delta$ is defined by the following relation
    \begin{align}
        E = I - \frac{R_\mathrm{Yb}}{(n-\delta)^2}, \label{eq:defect}
    \end{align}
    where $I = \SI{50443.07041}{cm^{-1}}$ and $R_\mathrm{Yb}=\SI{10973.696959}{cm^{-1}}$ are the energy of the first ionization limit and the Rydberg constant of ${}^{174}$Yb, respectively~\cite{Lehec:2018}. 
    The quantum defects for the low-lying ${}^3$D$_3$ state of $n=10$ and 12--15 were deduced in~\cite{Wyart:1979}.
    We confirm that the deduced quantum defects for the high-lying ${}^3$D$_3$ Rydberg states are in good agreement with the values for the low-lying ones. 
    Table \ref{table:rydberg} summarizes the energies and the quantum defects obtained by our measurement.
    
    \setlength{\tabcolsep}{12pt}
        \begin{table*}\centering
        \caption{\label{table:rydberg}Energy and quantum defect values of the observed resonances for $n=65$--80. Index numbers indicate the number assigned to each peak in Fig.\ref{fig:Rydberg_spectrum}(c). Typical uncertainty of the resonance frequency is about \SI{10}{MHz} coming from the uncertainty of the wavemeter.}
            \begin{tabular}{cccccccc}
                \toprule
                $n$  & index & $\delta$ & $E/h$ (MHz) & $n$  & index & $\delta$ & $E/h$ (MHz) \\\midrule
                65 & 3 & 2.753 118 00 & 1 511 396 147 & 74 & 6 & 2.732 881 65 & 1 511 597 475\\
                   & 4 & 2.749 201 30 & 1 511 396 253 &    & 7 & 2.728 919 37 & 1 511 597 547\\
                   & 7 & 2.728 954 75 & 1 511 396 805 &    & 8 & 2.713 498 12 & 1 511 597 827\\\hline
                66 & 3 & 2.753 233 55 & 1 511 422 780 & 75 & 1 & 2.757 558 00 & 1 511 614 846\\
                   & 4 & 2.749 123 37 & 1 511 422 887 &    & 3 & 2.754 396 71 & 1 511 614 902\\
                   & 7 & 2.729 019 84 & 1 511 423 410 &    & 4 & 2.751 617 38 & 1 511 614 950\\\cline{1-4}
                67 & 3 & 2.752 989 91 & 1 511 448 189 &    & 5 & 2.749 453 21 & 1 511 614 988\\
                   & 4 & 2.749 030 83 & 1 511 448 288 &    & 6 & 2.734 156 93 & 1 511 615 255\\
                   & 6 & 2.710 514 77 & 1 511 449 242 &    & 7 & 2.728 839 26 & 1 511 615 347\\
                   & 7 & 2.728 676 00 & 1 511 448 792 &    & 8 & 2.713 478 24 & 1 511 615 615\\\cline{5-8}
                   & 8 & 2.713 271 53 & 1 511 449 174 & 76 & 1 & 2.758 686 17 & 1 511 631 923\\\cline{1-4}
                68 & 3 & 2.753 556 14 & 1 511 472 419 &    & 2 & 2.756 319 39 & 1 511 631 963\\
                   & 4 & 2.749 545 81 & 1 511 472 514 &    & 3 & 2.754 128 63 & 1 511 631 999\\
                   & 7 & 2.729 072 10 & 1 511 472 999 &    & 4 & 2.751 338 22 & 1 511 632 046\\
                   & 8 & 2.713 037 07 & 1 511 473 378 &    & 5 & 2.748 931 06 & 1 511 632 086\\\cline{1-4}
                69 & 3 & 2.753 751 90 & 1 511 495 570 &    & 6 & 2.734 955 80 & 1 511 632 320\\
                   & 4 & 2.750 236 53 & 1 511 495 649 &    & 7 & 2.728 140 55 & 1 511 632 434\\
                   & 5 & 2.749 379 75 & 1 511 495 669 &    & 8 & 2.713 174 28 & 1 511 632 685\\\cline{5-8}
                   & 6 & 2.719 565 11 & 1 511 496 343 & 77 & 1 & 2.760 353 76 & 1 511 648 306\\
                   & 7 & 2.729 107 78 & 1 511 496 127 &    & 2 & 2.758 213 71 & 1 511 648 341\\
                   & 8 & 2.713 083 52 & 1 511 496 489 &    & 3 & 2.754 547 05 & 1 511 648 400\\\cline{1-4}
                70 & 3 & 2.753 698 45 & 1 511 517 700 &    & 4 & 2.751 787 62 & 1 511 648 444\\
                   & 4 & 2.750 364 04 & 1 511 517 772 &    & 5 & 2.749 265 15 & 1 511 648 485\\
                   & 5 & 2.749 266 63 & 1 511 517 796 &    & 6 & 2.736 433 65 & 1 511 648 691\\
                   & 6 & 2.722 625 17 & 1 511 518 372 &    & 7 & 2.728 692 66 & 1 511 648 815\\
                   & 7 & 2.728 931 74 & 1 511 518 236 &    & 8 & 2.713 606 30 & 1 511 649 057\\\cline{5-8}
                   & 8 & 2.713 374 54 & 1 511 518 572 & 78 & 1 & 2.762 192 34 & 1 511 664 039\\\cline{1-4}
                71 & 3 & 2.753 606 97 & 1 511 538 866 &    & 2 & 2.760 144 54 & 1 511 664 071\\
                   & 4 & 2.750 438 77 & 1 511 538 932 &    & 3 & 2.754 872 56 & 1 511 664 152\\
                   & 5 & 2.749 153 25 & 1 511 538 958 &    & 4 & 2.751 981 88 & 1 511 664 197\\
                   & 6 & 2.725 552 12 & 1 511 539 446 &    & 5 & 2.749 430 08 & 1 511 664 236\\
                   & 7 & 2.728 750 50 & 1 511 539 380 &    & 6 & 2.737 113 54 & 1 511 664 426\\
                   & 8 & 2.712 551 98 & 1 511 539 715 &    & 7 & 2.728 904 26 & 1 511 664 553\\\cline{1-4}
                72 & 3 & 2.753 973 35 & 1 511 559 112 &    & 8 & 2.713 498 64 & 1 511 664 791\\\cline{5-8}
                   & 4 & 2.750 944 03 & 1 511 559 172 & 79 & 1 & 2.764 389 69 & 1 511 679 153\\
                   & 5 & 2.749 409 48 & 1 511 559 203 &    & 3 & 2.755 316 70 & 1 511 679 287\\
                   & 6 & 2.728 897 38 & 1 511 559 609 &    & 4 & 2.752 270 80 & 1 511 679 333\\
                   & 7 & 2.728 880 20 & 1 511 559 609 &    & 5 & 2.749 658 77 & 1 511 679 371\\
                   & 8 & 2.713 575 35 & 1 511 559 912 &    & 6 & 2.737 673 32 & 1 511 679 549\\\cline{1-4}
                73 & 3 & 2.754 176 31 & 1 511 578 503 &    & 7 & 2.728 778 71 & 1 511 679 681\\
                   & 4 & 2.751 246 56 & 1 511 578 559 &    & 8 & 2.714 288 47 & 1 511 679 896\\\cline{5-8}
                   & 5 & 2.749 532 47 & 1 511 578 592 & 80 & 1 & 2.766 103 81 & 1 511 693 691\\
                   & 6 & 2.730 954 96 & 1 511 578 944 &    & 2 & 2.763 960 40 & 1 511 693 722\\
                   & 7 & 2.729 037 71 & 1 511 578 980 &    & 3 & 2.755 277 19 & 1 511 693 846\\
                   & 8 & 2.713 052 18 & 1 511 579 283 &    & 4 & 2.752 128 81 & 1 511 693 891\\\cline{1-4}
                74 & 1 & 2.756 543 85 & 1 511 597 045 &    & 5 & 2.749 345 23 & 1 511 693 930\\
                   & 3 & 2.754 300 03 & 1 511 597 085 &    & 6 & 2.737 524 41 & 1 511 694 099\\
                   & 4 & 2.751 552 23 & 1 511 597 135 &    & 7 & 2.728 632 77 & 1 511 694 226\\
                   & 5 & 2.749 510 64 & 1 511 597 172 &    & 8 & 2.713 481 34 & 1 511 694 442\\
                \bottomrule
            \end{tabular}
    \end{table*}

\section{Conclusion}
\label{sec:conclusion}
    In conclusion, we demonstrate core capabilities for quantum information processing based on single neutral Yb atoms ---  trapping and imaging of single atoms in optical tweezer arrays, the preparation of defect free arrays in 1D and 2D by atom rearrangement through the measurement and feedback, the direct optical excitation to the long-lived metastable state of ${}^3$P$_2$, and the demonstration of the single-photon excitation to Rydberg states within a tweezer array. 
    In addition, our results of Rydberg state spectroscopy show new observation of the ${}^3$D$_3$ series thanks to our choice of $J=2$ as the initial state.
    With the present work, the single qubit operation utilizing the rich internal degrees of freedom of nuclear spin and orbital degrees of freedom, two-qubit operation via a Rydberg state~\cite{Saffman:2010}, and the single site addressing with a magnetic field gradient~\cite{Shibata:2009}, are well within the reach.     

\section{Acknowledgement}
\label{sec:acknowledgement}
We thank Abhilash Kumar Jha, Phillipp Lunt, Vikram Ramesh, and Kota Yamamoto for the experimental assistance. DO acknowledges support from the JSPS (KAKENHI grant number JP21J11878). This work was supported by the Grant-in-Aid for Scientiﬁc Research of JSPS(Nos. JP17H06138, JP18H05405, JP18H05228, and JP21H01014), the Impulsing Paradigm Change through Disruptive Technologies (ImPACT) program, JST CREST (No. JP-MJCR1673), and MEXT Quantum Leap Flagship Program (MEXT Q-LEAP) Grant No. JPMXS0118069021.

\bibliographystyle{apsrev4-2}
\bibliography{a}

\end{document}